\documentclass{amsart}

\usepackage{times}
\usepackage{amssymb,amsmath,amsfonts}
\usepackage{amsthm}
\usepackage{latexsym}
\usepackage{enumerate}

\usepackage{amssymb}

\newcommand{\Prof}{\ensuremath{\mathcal{P}}}

\newcommand{\Leaves}{\ensuremath{\mathcal{L}}}

\newtheorem{theorem}{Theorem}
\newtheorem{lemma}{Lemma}

\theoremstyle{remark}

\theoremstyle{definition}

\begin{document}

\title{Graph Triangulations and the Compatibility of Unrooted Phylogenetic Trees}

\author[S. Vakati]{Sudheer Vakati}

\address{Department of Computer Science, Iowa State University, Ames, Iowa 50011, U.S.A.}
\email{svakati@iastate.edu}

\author[D. Fern\'{a}ndez-Baca]{David Fern\'{a}ndez-Baca}

\address{Department of Computer Science, Iowa State University, Ames, Iowa 50011, U.S.A.}

\email{fernande@cs.iastate.edu}

\keywords{Compatibility, chordal graphs, graph triangulation, phylogenetics, supertrees, tree decompositions}

\begin{abstract}
We characterize the compatibility of a collection of unrooted phylogenetic trees as a question of determining whether a graph derived from these trees --- the display graph --- has a specific kind of triangulation, which we call legal.   Our result is a counterpart to the well known triangulation-based characterization of the compatibility of undirected multi-state characters.

\end{abstract}

\maketitle

\section{Introduction}

A  \emph{phylogenetic tree} or \emph{phylogeny} is an unrooted tree $T$ whose leaves are in one-to-one correspondence with a set of \emph{labels} (\emph{taxa})
$\Leaves(T)$.  If $\Leaves(T) = X$, we say that $T$ is a \emph{phylogenetic tree for $X$}, or a \emph{phylogenetic $X$-tree} \cite{SempleSteel03}.  A phylogenetic tree represents the evolutionary history of a set of species, which are the labels of the tree.

Suppose $T$ is a phylogenetic tree.  Given a subset $Y \subseteq \Leaves(T )$, the \emph{subtree of
$T$ induced by $Y$}, denoted $T |Y$, is the tree obtained by forming the
minimal subgraph of $T$ connecting the leaves with labels in $Y$ and then suppressing vertices of degree two.  Let $T'$ be some other phylogenetic tree such that $\Leaves(T') \subseteq \Leaves(T)$.  We say that $T$ \emph{displays} $T'$ if $T'$ can be obtained by contracting
edges in the subtree of $T$ induced by $\Leaves(T')$.

A \emph{profile} is a tuple $\Prof = (T_1, T_2, \dots , T_k)$, where each $T_i$ is a phylogenetic tree for some set of labels $\Leaves(T_i)$.  The $T_i$s are called \emph{input trees}, and we may have $\Leaves(T_i) \cap \Leaves(T_j) \neq \emptyset$ for $i \ne j$.  A \emph{supertree} for $\Prof$ is a phylogeny $T$ with $\Leaves(T) = \bigcup_{i=1}^k \Leaves(T_i)$.  Profile $\Prof$ is \emph{compatible} if
there exists a supertree $T$ for $\Prof$ that displays $T_i$, for each $i \in \{1, \ldots , k\}$. The \emph{phylogenetic tree compatibility problem} asks, given a profile $\Prof$, whether or not $\Prof$ is compatible.  This question arises when trying to assemble a collection of phylogenies for different sets of species into a single phylogeny (a supertree) for all the species~\cite{Gordon86}.  The phylogenetic tree compatibility problem asks whether or not it is possible to do so via a supertree that does not conflict with any input tree.

Phylogenetic tree compatibility is NP-complete~\cite{Steel92} (but the problem is polynomially-solvable for rooted trees~\cite{ASSU81}).  Nevertheless, Bryant and Lagergren have shown that the problem is fixed-parameter tractable for fixed $k$ \cite {BryantLagergren06}.  Their argument relies on a partial characterization of compatibility in terms of tree-decompositions and tree-width of a structure that they call the ``display graph'' of a profile (this graph is defined in Section~\ref{sec:LT}).  Here we build on their argument to produce a complete characterization of compatibility in terms of the existence of a special kind of triangulation of the display graph.  These \emph{legal} triangulations (defined in Section~\ref{sec:LT}) only allow certain kinds of edges to be added.  Our result is a counterpart to the well-known characterization of character compatibility in terms of triangulations of a class of intersection graphs~\cite{Bun74}, which has algorithmic consequences~\cite{Gusfield:2009, McMorrisWarnowWimer94}. Our characterization of tree compatibility may have analogous implications.

\section{Preliminaries}

Let $G$ be a graph.  We write $V(G)$ and $E(G)$ to denote the vertex set and edge set of $G$, respectively.  Suppose $C$ is a cycle in $G$.   A \emph{chord} in $C$ is any edge of $G$ whose endpoints are two nodes that are not adjacent in $C$.  $G$ is said to be \emph{chordal} if and only if it every cycle of length at least four has a chord. A graph $G'$ is a \emph{chordal fill-in} or \emph{triangulation} of $G$ if $V(G') =V(G)$, $E(G') \supseteq E(G)$, and $G'$ is chordal. The set $E(G')  \setminus E(G)$ is called a \emph{fill-in} for $G$ and the edges in it are called \emph{fill-in edges}. 

A \emph{tree decomposition} for a graph $G$ is a pair $(T,B)$, where $T$ is a tree and $B$ is a mapping from $V (T)$ to subsets of $V (G)$ that satisfies the following three properties.
\begin{enumerate}[(TD1)]
\item (\emph{Vertex Coverage}) For every $v \in V (G)$ there is an $x \in V (T )$ such that $v \in B(x)$.
\item (\emph{Edge Coverage}) For every edge $\{u, v\} \in E(G)$ there exists an $x \in V (T )$ such that $\{u, v\} \subseteq B(x)$.
\item (\emph{Coherence})
For every $u \in V (G)$ the set of vertices $\{x \in V(T) : u \in B(x)\}$ forms a subtree of $T$.
\end{enumerate}

It is well known that if $G$ is chordal, $G$ has a tree-decomposition $(T,B)$ where (i) there is a one-to-one mapping $C$ from the vertices of $T$ to the maximal cliques of $G$ and (ii) for each vertex $x$ in $T$, $B(x)$ consists precisely of the vertices in the clique $C(x)$~\cite{Heggernes:2005}.  This sort of tree decomposition is called a \emph{clique tree} for $G$.   Conversely, let $(T,B)$ be  a tree decomposition of a graph $G$ and let $F$ be the set of all $\{u,v\} \notin E(G)$ such that $\{u, v\}  \subseteq B(x)$ for some $x \in V(T)$.  Then, $F$ is a chordal fill-in for $G$~\cite{Heggernes:2005}.
We shall refer to this set $F$ as the \emph{chordal fill-in of $G$ associated with tree-decomposition $(T,B)$} and to the graph $G'$ obtained by adding the edges of $F$ to $G$ as the \emph{triangulation of $G$ associated with $(T,B)$}.

\section{Legal Triangulations and Compatibility}

\label{sec:LT}

The \emph{display graph} of a profile $\Prof = (T_1, \dots , T_k)$ is the graph $G = G(\Prof)$ formed from the disjoint graph union of $T_1, \dots , T_k$ by identifying the leaves with common labels (see Fig. 1 of~\cite{BryantLagergren06}).
An  edge $e$ of $G$ is \emph{internal} if, in the input tree where it originated, both
endpoints of $e$ were internal vertices; otherwise, $e$ is \emph{non internal}.  A vertex $v$ of $G$ is called a \emph{leaf} if it was obtained by identifying input tree leaf nodes with the same label $\ell$.   The label of $v$ is $\ell$.   A non-leaf vertex of $G$ is said to be \emph{internal}.

A triangulation $G'$ of the display graph $G$ is \emph{legal} if it satisfies the following conditions.
\begin{enumerate}[(LT1)]
\item
Suppose a clique in $G'$ contains an internal edge.  Then, this clique can contain no other edge from $G$ (internal or non internal).

\item
Fill-in edges can only have internal vertices as their endpoints.
\end{enumerate}

Note that the above conditions rule out a chord between vertices of
the same tree. Also, in any legal triangulation of $G$, any clique that contains
a non internal edge cannot contain an internal edge from any tree.

The importance of legal triangulations derives from the next results, which are proved in the next section.

\begin{lemma}
\label{lm:dir2} 
Suppose a profile $\Prof = (T_1, \dots, T_k)$ of unrooted phylogenetic trees is compatible.  Then the display graph of $\Prof$ has a legal triangulation.
\end{lemma}

\begin{lemma}
\label{lm:dir1}
Suppose the display graph of a profile $\Prof =  (T_1, \dots , T_n)$ of unrooted trees has a legal triangulation.  Then $\Prof$ is compatible.
\end{lemma}

The preceding lemmas immediately imply our main result.

\begin{theorem} 
\label{thm:main}
A profile $\Prof = (T_1, \dots, T_k)$ of unrooted trees is
compatible if and only if the display graph of $\Prof$ has a
legal triangulation.
\end{theorem}

\section{Proofs}

The proofs of Lemmas~\ref{lm:dir2} and~\ref{lm:dir1} rely on a new concept.  Suppose $T_1$ and $T_2$ are phylogenetic trees such that $\Leaves(T_2) \subseteq \Leaves(T_1)$.  An \emph{embedding function from $T_1$ to $T_2$} is a surjective map $\phi$ from a subgraph of $T_1$ to $T_2$ satisfying the following properties.
\begin{enumerate}[(EF1)]  
\item $\phi$ maps labeled vertices to vertices with the same label.
\item For every vertex $v$ of $T_2$ the set  $\phi^{-1}(v)$ is a connected subgraph of $T_1$.
\item
For every edge $\{u, v\}$ of $T_2$ there is a unique edge $\{u', v'\}$ in $T_1$ such that $\phi(u') = u$ and $\phi(v') = v$.
\end{enumerate}

The next result extends Lemma 1 of ~\cite{BryantLagergren06}.

\begin{lemma}
Let $T_1$ and $T_2$ be phylogenetic trees and $\Leaves(T_2) \subseteq \Leaves(T_1)$. Tree $T_1$ displays Tree $T_2$ if and only if there exists an embedding function $\phi$ from $T_1$ to $T_2$.
\end{lemma}

\begin{proof}
The ``only if'' part was already observed by Bryant and Lagergren (see Lemma 1 of ~\cite{BryantLagergren06}).  We now prove the other direction.
 
To prove that $T_1$ displays $T_2$, we argue that $T_2$ can be obtained from $T_1|\Leaves(T_2)$ by a series of edge contractions, which are determined by the embedding function $\phi$ from $T_1$ to $T_2$.  
Let $T_1'$ be the graph obtained from $T_1|\Leaves(T_2)$ by considering each  vertex $v$ of $T_2$ and identifying all vertices of $\phi^{-1}(v)$ in $T_1|\Leaves(T_2)$ to obtain a single vertex $u'$ with $\phi(u') = v$.  By property (EF2), each such step yields a tree.  By properties (EF1)--(EF3), each vertex $v$ of $T_1|\Leaves(T_2)$ is in the domain of $\phi$.  Thus, function $\phi$ is now a bijection between $T_2$ and $T_1'$ that satisfies (EF1)--(EF3).

We claim that for any two vertices $u, v \in V(T_2)$, there is an edge $\{u, v\} \in E(T_2)$ if and only if there is an edge $\{\phi^{-1}(u), \phi^{-1}(v)\} \in E(T_1')$. The ``only if'' part follows from property (EF3). For the other direction, assume by way of contradiction that $\{x, y\} \notin E(T_2)$, but that $\{\phi^{-1}(x), \phi^{-1}(y)\} \in E(T_1')$. Let $P$ be the path between vertices $x$ and $y$ in $T_2$.  By property (EF3), there is a path between nodes $\phi^{-1}(x)$, $\phi^{-1}(y)$ in tree $T_1'$
that does not include the edge $\{\phi^{-1}(x), \phi^{-1}(y)\}$. This path along with the edge $\{\phi^{-1}(x), \phi^{-1}(y)\}$ forms a cycle in $T_1'$, which gives the desired contradiction.

Thus, the bijection $\phi$ between $T_2$ and $T_1'$ is actually an isomorphism between the two trees. It now follows from property (EF1) that $T_1$ displays $T_2$.
\end{proof}

The preceding lemma immediately implies the following characterization of compatibility. 

\begin{lemma}\label{lm:BLS}
Profile $\Prof = (T_1, \dots , T_k)$ is compatible if and only if there exist a supertree $T$ for $\Prof$ and functions $\phi_1, \dots , \phi_k$, where, for $i = 1, \dots, k$, $\phi_i$ is an embedding function from $T$
to $T_i$.
\end{lemma}

\begin{proof} [Proof of Lemma~\ref{lm:dir2}]
If $\Prof$ is compatible, there exists a supertree for $\Prof$ that displays $T_i$ for $i = 1, \dots , k$.  Let $T$ be any such supertree.   By Lemma~\ref{lm:BLS}, for $i = 1, \dots , k$, there exists an embedding function $\phi_i$ from $T$ to $T_i$.  We will use $T$ and the $\phi_i$s to build a tree decomposition $(T_G, B)$ corresponding to a legal triangulation $G'$ of the display graph $G$ of~$\Prof$.  The construction closely follows that given by 
Bryant and Lagergren in their proof of Theorem 1 of~\cite{BryantLagergren06}; thus, we only summarize the main ideas.   

Initially we set $T_G = T$ and, for every $v \in V(T)$,
$B(v) = \{\phi_i(v) : v \text{ in the domain of } \phi_i; 1 \le i \le k\}$.
Now, $(T_G,B)$ satisfies the vertex coverage property and the coherence property, but not edge coverage \cite{BryantLagergren06}.
To obtain a pair $(T_G,B)$ that satisfies all three properties, subdivide the edges of $T_G$ and extend $B$ to the new vertices.  Do the following for each edge $\{x,y\}$ of $T_G$.  Let $F = \{\{u_1, v_1\}, \dots , \{u_m, v_m\}\}$ be set of edges of $G$ such that $u_i \in B(x)$ and $v_i \in B(y)$.  Observe that $F$ contains at most one edge from $T_i$, for $i = 1, \dots, k$  (thus, $m \leq k$).  Replace edge $\{x, y\}$ by a path $x, z_1, \dots , z_m, y$, where $z_1, \dots , z_m$ are new vertices. For $i = 1, 2, \dots , m$, let
$B(z_i ) = (B(x) \cap B(y)) \cup \{v_1, \dots , v_i, u_i, \dots , u_m\}$.
The resulting pair $(T_G, B)$ can be shown to be a tree decomposition of $G$ of width $k$ (see~\cite{BryantLagergren06}).

The preceding construction guarantees that $(T_G,B)$ satisfies two additional properties:
\begin{enumerate}[(i)]
\item
For any $x \in V(T_G)$, if $B(x)$ contains both endpoints of an internal edge of $T_i$, for some $i$, then $B(x)$ cannot contain both endpoints of any other edge, internal or not.
\item
Let $x \in V(T_G)$ be such that $B(x)$ contains a labeled vertex $v \in V(G)$.  Then, for every $u \in B(x) \setminus \{v\}$, $\{v,u\} \in E(G)$.
\end{enumerate}

Properties (i) and (ii) imply that the triangulation of $G$ associated with $(T_G,B)$ is legal.
\end{proof}

Next, we prove Lemma~\ref{lm:dir1}.  For this, we need some definitions and auxiliary results. Assume that the display graph of profile $\Prof$ has a legal triangulation $G'$.  Let $(T', B)$ be a clique tree for $G'$. For each vertex $v \in V(G)$, let $N(v)$ denote the set of all nodes in the clique tree $T'$ that contain $v$. Observe that the coherence property implies that $N(v)$ induces a subtree of $T'$.

\begin{lemma}
\label{claim1}
Suppose vertex $v$ is a leaf in tree $T_i$, for some $i \in \{1, \dots,k\}$.  Let $U(v) = \bigcup_{x \in N(v)} B(x)$.  Then, for any $j \in \{1, \dots,k\}$, 
at most one internal vertex $u$ from input tree $T_j$ is
present in $U(v)$.  Furthermore, for any such a vertex $u$ we must have that $\{u,v\} \in E(G)$. 
\end{lemma}

\begin{proof}
Follows from condition (LT2).
\end{proof}

\begin{lemma}
\label{claim2}
Suppose $e = \{u,v\}$ is an internal edge from input tree $T_i$, for some $i \in \{1, \dots,k\}$.  Let $U(e) = \bigcup_{x \in \{N(u) \cap N(v)\}}B(x)$.  Then, 
\begin{enumerate}[(i)]
\item
$U(e)$ contains at most one vertex of $T_j$, 
for any $j \in \{1, \dots, k\}$, $j \ne i$, and 
\item
$V(T_i) \cap U(e) = \{u, v\}$. 
\end{enumerate}
\end{lemma}

\begin{proof}
Part (ii) follows from condition (LT1). We now prove part (i).

Assume by way of contradiction that the claim is false.  Then,
there exists a $j \neq i$ and an edge $\{x, y\} \in T'$ such that $e \subseteq B(x)$, $e \subseteq B(y)$, and there are vertices $a, b \in V(T_j)$, $a \neq b$, such that $a \in B(x)$ and $b \in B(y)$.

Deletion of edge $\{x, y\}$ partitions $V(T')$ into two sets $X$ and $Y$.  Let $P = \{a \in V(T_j) : a \in B(z) \text{ for some } z \in X\}$ and $Q = \{b \in V(T_j) : b \in B(z) \text{ for some } z \in Y\}$.
By the coherence property, $(P,Q)$ is a partition of $V(T_j)$. There must be a vertex $p$ 
in set $P$ and a vertex $q$ in set $Q$ such that $\{p, q\} \in E(T_j)$. Since $G'$ is a 
legal triangulation, there must be a node $z$ in $T'$ such that
$p, q \in B(z)$. Irrespective of whether $z$ is in set $X$ or $Y$, 
the coherence property is violated, a contradiction.
\end{proof}

A legal triangulation of the display graph of a profile is \emph{concise} if 
\begin{enumerate}[(C1)]
\item
each internal edge is contained in exactly one maximal clique in the triangulation and
\item
every vertex that is a leaf in some tree is contained in exactly one maximal clique of the triangulation.
\end{enumerate}

\begin{lemma}
\label{lm:concise}
Let $G$ be the display graph of a profile $\Prof$.  If $G$ has a legal triangulation, then $G$ has a concise legal triangulation.
\end{lemma}

\begin{proof}
Let $G'$ be a legal triangulation of the display graph $G$ of profile $\Prof$ that is not concise.  Let $(T',B)$ be a clique tree for $G'$.  
We will build a concise legal triangulation for $G$ by repeatedly applying contraction operations on $(T',B)$. 
The \emph{contraction} of an edge $e = \{x,y\}$ in $T'$ is the operation that consists of (i) replacing $x$ and $y$ by a single (new) node $z$, (ii) adding edges from node $z$ to every neighbor of $x$ and $y$,
and (iii) making $B(z) = B(x) \cup B(y)$. Note that the resulting pair $(T',B)$ is a tree decomposition for $G$ (and $G'$); however, it is not guaranteed to be a clique tree for $G'$.

We proceed in two steps.  First, for every leaf $v$ of $G$ such that $|N(v)| > 1$, contract each edge $e = \{x,y\}$ in $T'$ such that $x, y \in N(v)$.
In the second step, we consider each edge $e = \{u,v\}$ of $G$ such that $|N(u) \cap N(v)| > 1$, contract each edge $\{x,y\}$ in $T'$ such that $x, y \in N(u) \cap N(v)$.
Lemma~\ref{claim1} (respectively, Lemma~\ref{claim2}) ensures that each contraction done in the first (respectively, second) step leaves us with a new tree decomposition whose associated triangulation is legal.
Furthermore, the triangulation associated with the final tree decomposition is concise. 
\end{proof}

\begin{proof}[Proof of Lemma~\ref{lm:dir1}]
We will show that,  given a  legal triangulation $G'$ of  the display graph $G$ of profile $\Prof$, we can generate a supertree $T$ for $\Prof$ along with an embedding function $\phi_i$ from $T$ to $T_i$, for $i = 1, \dots, k$.  By Lemma~\ref{lm:BLS}, this immediately implies that $\Prof$ is compatible 

By Lemma~\ref{lm:concise}, we can assume that $G'$ is concise. Let $(T',B)$ be a clique tree for $G'$.  Initially, we make $T = T'$.
Next, for each node $x$ of $T$, we consider three possibilities:
\begin{enumerate}[{Case} 1:]
\item
\emph{$B(x)$ contains a labeled vertex $v$ of $G$.}  Then, $v$ is a leaf in some input tree $T_i$; further, by conciseness, $x$ is the unique node in $T$ such that $v \in B(x)$, and, by the edge coverage property, if $u$ is the neighbor of $v$ in $T_i$, $u\in B(x)$.  Now, do the following.
\begin{enumerate}[(i)]
\item
Add a new node $x_v$ and a new edge $\{x, x_v\}$  to $T$.  
\item
Label $x_v$ with $\ell$, where $\ell$ is the label of $v$. 
\item
For each $i \in \{1, \dots, k\}$ such that $v$ is a leaf in $T_i$, make $\phi_i(x_v) = v$ and $\phi_i(x) = u$, where $u$ is the neighbor of $v$ in $T_i$. 
\end{enumerate}

\item
\emph{$B(x)$ contains both endpoints of an internal edge $e$ of some input tree $T_i$.}  By legality, $B(x)$ does not contain both endpoints of any other edge of any input tree, and, by conciseness, $x$ is the only node of $T$ that contains both endpoints of $e$.   Now, do the following.
\begin{enumerate}[(i)]
\item
Replace node $x$ with nodes $x_u$ and $x_v$, and add edge $\{x_u, x_v\}$.  
\item
Add an edge between node $x_u$ and every node neighbor $y$ of $x$ such that $u \in B(y)$. 
\item
Add an edge between node $x_v$ and every neighbor $y$ of $x$ such that $v \in B(y)$. 
\item For each neighbor $y$ of $x$ such that $u \notin B(y)$ and $v \notin B(y)$, add an edge from $y$ to node $x_u$ or node $x_v$, but not to both (the choice of which edge to add is arbitrary).  
\item
Make $\phi_i(x_u) = u$ and $\phi_i(x_v) =v$.
\end{enumerate}

\item
\emph{$B(x)$ contains at most one internal vertex from $T_i$ for $i \in \{1, \dots, k\}$.}  Then,  for every $i$ such that $B(x) \cap V(T_i) \ne \emptyset$ make $\phi_i(x) = v$, where $v$ is the vertex of $T_i$ contained in $B(x)$.  
\end{enumerate}

By construction (Case 1) and the legality and conciseness of $(T',B)$, for every $\ell \in \bigcup_{i=1}^k \Leaves(T_i)$ there is exactly one leaf $x \in V(T)$ that is labeled $\ell$.
Thus, $T$ is a supertree of profile $\Prof$.  Legality also ensures that the function $\phi_i$ is a surjective map from a subgraph of $T$ to $T_i$.  Furthermore, the handling of Case 1 guarantees that $\phi_i$ satisfies (EF1). The coherence of $(T',B)$ and ensures that $\phi_i$ satisfies (EF2). The handling of Case 2 and conciseness ensure that $\phi_i$ satisfies (EF3). Thus, $\phi_i$ is an embedding function, and, by Lemma~\ref{lm:BLS}, profile $\Prof$ is compatible.
\end{proof}

\section*{Acknowledgements}
This work was supported in part by the National Science Foundation under grants DEB-0334832 and DEB-0829674.

%

\end{document}